\documentclass[aps,pra,twocolumn,showpacs,10pt,amsmath,footinbib]{revtex4-1}
\usepackage{graphicx}
\usepackage[usenames,dvipsnames]{xcolor}
\usepackage{braket}
\usepackage{array}
\newcolumntype{x}[1]{>{\centering\let\newline\\\arraybackslash\hspace{0pt}}p{#1}}
\newcommand{\pa}{\partial}

\newcommand{\salt}{\,\mbox{\footnotesize$\lesssim$}\,}

\usepackage{color}

\begin{document}

\title{Critical velocity for vortex nucleation in a finite-temperature Bose gas}

\author{G. W. Stagg}
\affiliation{Joint Quantum Centre (JQC) Durham--Newcastle, School of Mathematics and Statistics, Newcastle University, Newcastle upon Tyne, NE1 7RU, United Kingdom}
\author{R. W. Pattinson}
\affiliation{Joint Quantum Centre (JQC) Durham--Newcastle, School of Mathematics and Statistics, Newcastle University, Newcastle upon Tyne, NE1 7RU, United Kingdom}
\author{C. F. Barenghi}
\affiliation{Joint Quantum Centre (JQC) Durham--Newcastle, School of Mathematics and Statistics, Newcastle University, Newcastle upon Tyne, NE1 7RU, United Kingdom}
\author{N. G. Parker}\email{nick.parker@ncl.ac.uk}
\affiliation{Joint Quantum Centre (JQC) Durham--Newcastle, School of Mathematics and Statistics, Newcastle University, Newcastle upon Tyne, NE1 7RU, United Kingdom}

\date{\today}

\begin{abstract}
We use classical field simulations of the homogeneous Bose gas to study
the breakdown of superflow due to vortex nucleation past a cylindrical obstacle at finite temperature.
 Thermal fluctuations modify the vortex nucleation from the obstacle,
turning anti-parallel vortex lines (which would be nucleated at zero
temperature) into wiggly lines, vortex rings and even vortex tangles. We find that
the critical velocity for vortex nucleation decreases with
increasing temperature, and scales with the speed of sound of
the condensate, becoming zero at the critical temperature for condensation.
\end{abstract}

\pacs{03.75.Kk, 03.75.Mn, 05.65.+b, 67.85.Fg}

\maketitle

  \section{Introduction}
A defining feature of superfluids is the absence of excitations when the flow (relative to some obstacle or boundary) is slower than a critical velocity; above this velocity, the flow becomes dissipative.  This can be understood in terms of the Landau criterion, which predicts excitations when the local fluid velocity exceeds $v_{\rm L} = \textrm{min} [E(p)/p]$, where $p$ is the momentum of elementary excitations and $E(p)$ their energy \cite{NozieresPines}.  In weakly-interacting atomic Bose-Einstein
condensates, and for infinitesimally small perturbations, one obtains $v_{\rm L} = c$, the speed of sound.  The breakdown of superfluidity has been experimentally probed by introducing a localized repulsive obstacle, engineered via the repulsive force generated by focussed blue-detuned laser beam, and moving the condensate relative to the obstacle \cite{Neely,kwon_moon_14,kwon_2015a,kwon_2015b,Raman,Onofrio,Inouye,desbuquois_2012}.  This has enabled measurement of the critical velocity and the direct observation of the ensuing excitations, that is, pairs of quantized vortex lines with opposite polarity.
In flattened condensates, this scenario currently provides
a route to engineer states of two-dimensional quantum turbulence
\cite{Neely,kwon_moon_14}; it also gives
insight into the deep link between quantum fluids and their classical
counterparts, where it has been predicted that the wake of quantized
vortices produced downstream of the obstacle can collectively
mimick the classical wakes, including
the B{\' e}rnard-von K{\'a}rm{\' a}n vortex street
\cite{saito_2010,stagg_2014,reeves_2015}.

The motion of an obstacle in the zero-temperature Bose gas, described by the Gross-Pitaevskii equation, is a well-studied problem, particularly for circular obstacles in 2D geometries.  The pioneering simulations by Frisch {\it et al.} \cite{frisch92} of an impenetrable circular obstacle moving within the 2D nonlinear Schr\"odinger equation (NLSE),
demonstrated the existence of a critical velocity of value
$v_{\rm c}\sim 0.4c$ above which vortex-antivortex pairs are nucleated.
For small obstacles, boundary effects tend to suppress vortex nucleation,
and, as the obstacle's size increases,
the critical velocity reduces towards an asymptotic value \cite{berloff_2000,rica_2001,pham_2004}.  The critical velocity also depends on the shape of the obstacle, for example, obstacles with elliptical cross-section lead to reduced/heightened $v_{\rm c}$, depending on the orientation relative to the flow \cite{stagg_2014, stagg_2015b}. Similar behaviour holds for spherical obstacles, albeit with the emission of vortex rings and increased critical speeds of circa $0.7 c$ \cite{winiecki_2000,win01,stagg_2014}.
In current condensate experiments \cite{Neely,kwon_moon_14,Raman,Onofrio,Inouye,desbuquois_2012, kwon_2015a,kwon_2015b}, the obstacles are penetrable, corresponding to a Gaussian potential of finite amplitude, produced via an incident blue-detuned laser beam.
The same qualitative behaviour emerges for impenetrable obstacles,
although the critical velocity and vortex nucleation patterns become modified \cite{saito_2010}.

Very recently, Kwon {\it et al.} have undertaken
a systematic experimental analysis of the critical velocity for vortex
shedding, exploring the dependence of the nucleation
on height and width of the penetrable obstacle and the crossover
from penetrable to impenetrable obstacles \cite{kwon_2015a}.
Their results, obtained in a condensate with temperature much lower than
the critical temperature for condensation, are
in agreement with previous zero-temperature predictions based on the
Gross-Pitaesvkii equation.
Their work has made a significant step in consolidating our
theoretical and experimental understanding of the critical velocity
in a condensate in the zero-temperature limit.  At the same time,
it has highlighted the need to extend the study
of the critical velocity to finite temperatures.
While the role of finite temperature has been explored considerably for another vortex nucleation scenario, namely within deformed, rotating traps \cite{hodby_2002,abo_shaeer_2002,williams_2002,penckwitt_2002,kasamatsu_2003,lobo_2004} (for which unstable surface modes underpin the vortex nucleation), there is a paucity of literature relating to the finite temperature behaviour of vortex nucleation by a translating obstacle.  Indeed, to our knowledge, the only finite-temperature analysis of a moving obstacle in a three-dimensional condensate is that of Leadbeater {\it et al.}
\cite{leadbeater_2003},
who found that the critical velocity of a hard sphere decreases
with temperature.

In this work we study the motion of a cylindrical Gaussian-shaped
obstacle through a three-dimensional homogeneous Bose gas at finite
temperature via classical field simulations. We
find that the critical velocity decreases with temperature and
increases with condensate fraction (ratio of condensate to total density).
Indeed, the critical velocity is found to be closely proportional
to the speed of sound of the condensate, which scales as the square
root of the condensate fraction. Above
the critical velocity, vortex nucleation occurs either through
pairs of vortex lines, collections of vortex rings, or direct formation of a vortex tangle, and we indicate the occurrence of these structures in the parameter space of condensate fraction and flow speed.

  \section{Classical Field Method}
\label{sec:theory}

We consider a weakly-interacting Bose gas with $N$ atoms in a periodic box of volume $\ell^3$.  The atoms have mass $m$ and their interactions are approximated by a contact pseudo-potential $V_{\mathrm{int}}({\bf r}-{\bf r'})= g \delta({\bf r}-{\bf r'})$, where $g$ is a coefficient which characterises the atomic interactions and $\delta$ is the Dirac delta function \cite{Pethick}.

In order to theoretically model thermal excitations of the weakly-interacting
Bose gas, one must progress beyond the standard mean-field approximation to
include both the condensate and the thermal fraction atoms in the gas.
Various methods have been proposed
for this purpose, as reviewed elsewhere
\cite{Pol_Rev,Proukakis,finite_temp_book2,finite_temp_book,Blakie,berloff_2014}.
Among these methods, a popular one
is the classical field method ~\cite{Svis5,Davis,PRL.87.210404,
PhysRevA.66.013603,Davis2,PhysRevLett.95.263901,Pol_Rev}.
This method is based on the observation that, providing the modes of
the gas are highly occupied (an {\it a priori} assumption in our work), then the gas can be
approximated by a classical field $\psi({\bf r},t)$
whose equation of motion is the Gross-Pitaevskii equation (GPE).
However, whereas the GPE
conventionally describes the condensate only, $\psi({\bf r},t)$
now describes the entire multi-mode `classical' gas \cite{Proukakis,Blakie}.  The classical field method has been used to model phenomena
beyond-mean-field effects, including thermal equilibration dynamics~\cite{PhysRevA.66.013603,PhysRevLett.95.263901,pattinson_2014,nazarenko_2014}, condensate fractions \cite{Davis}, critical temperatures \cite{Davis2006}, correlation functions \cite{Wright2011}, spontaneous production of vortex-antivortex pairs in quasi-2D gases \cite{Simula}, thermal dissipation of vortices \cite{berloff_2007},  and related effects in binary condensates \cite{Berloff_2006,Salman20091482,pattinson_2014}.

We parameterize the gas by
the classical field $\psi({\bf r},t)$.  The density distribution of atoms
is then $|\psi({\bf r},t)|^2$. The evolution
of $\psi$ is governed by the GPE
  \begin{eqnarray}
i  \hbar \frac{\pa\psi }{\pa t}=\left( -\frac{\hbar^2}{2m}\nabla^{2} + V_{\mathrm{obj}}({\bf r},t) +g\left|\psi \right|^{2} \right ) \psi,\label{gpe}
  \end{eqnarray}
where $V_{\mathrm{obj}}({\bf r},t)$ is the externally applied potential.
The GPE conserves the total number of particles,
$N = \int |\psi|^2~{\rm d}V$, and the total energy,
\[H=\int \left(\frac{\hbar^2}{2m}|\nabla \psi|^2 + V_{\mathrm{obj}}|\psi|^2+ \frac{g}{2}|\psi|^4  \right)~{\rm d}V.\]
In what follows we express all quantities in terms of the
natural units of the homogeneous Bose gas:  density in terms of a uniform value $\rho$, length in terms of the healing length $\xi=\hbar/\sqrt{m g \rho}$, speed in terms of the speed of sound $c=\sqrt{\rho g/m}$, energy in terms of the chemical potential of the homogeneous condensate $\mu=\rho g$, and time in terms of $\tau=\hbar / g \rho$.

We label the modes of the system through the wavevector ${\bf k}$.  To allow for occupation across all classical modes of the system, the initial condition is highly non-equilibrium,
  \begin{equation}
    \psi \left(\mathbf{r},0\right)=\sum_{\mathbf{k}}a_{\mathbf k}\exp(i\mathbf{k}\cdot\mathbf{r})
    \label{eq:rand2}
  \end{equation}
where the coefficients $a_{\mathbf k}$ are uniform and the phases
are distributed randomly~\cite{PhysRevA.66.013603}.   The occupation of mode ${\bf k}$ is
$n_{\mathbf{k}}=|a_{\mathbf{k}}|^2$. The final temperature/condensate fraction of our simulations is varied through a rescaling of $\psi$, so as to fix the quantities $N$ and $H$.

The GPE is evolved numerically, in the absence of any potential $V_{\mathrm{obj}}$, using a fourth-order Runge-Kutta method on a $192^3$ periodic grid with time step $\Delta t =0.01 \tau$ and isotropic grid spacing $\Delta =0.75\xi$. The spatial discretization of our numerical grid implies that high momenta are not described in our simulations. In effect, an ultraviolet cutoff is introduced, $n_{\mathbf{k}}(t)=0$ for $k>k_{\rm{max}}$, where $k=|{\bf k}|$ and the maximum described wave vector amplitude is $k_{\rm{max}} = \sqrt{3} \pi / \Delta$. 

The ensuing evolution from the strongly nonequilibrium initial conditions has been outlined previously \cite{PhysRevA.66.013603,pattinson_2014}.  Initially the mode occupation numbers $n_k$ are uniformly distributed over wavenumber $k$, up to the cutoff.  Self-ordering leads to the rapid growth in the occupation of low-$k$ modes, which initially evolves in a state of weak turbulence.  Then the distribution evolves to a bimodal form.
The high-$k$ part of the distribution is associated with the
thermal excitations and low mode occupations.
The low-$k$ part of the field is the quasi-condensate,
characterised by macroscopic mode populations and superfluid ordering.

From the bimodal distribution, a wavenumber $k_c$ can be chosen as the boundary in $k$-space between the quasi-condensate and the thermal gas, as performed in \cite{PhysRevA.66.013603}.  Here we take $k_c \approx 13~(2\pi N^{-1} \xi^{-1})$, although our qualitative results are insensitive to the precise definition of $k_{\rm c}$.  The condensate density, $\rho_0$, is then calculated as the density within the quasi-condensate, i.e. a coarse-grained averaging over the quasi-condensate modes.  This is then used to define the condensate fraction, $\rho_0/\rho$, where $\rho$ is the total density of the gas.

While the raw wavefunction is too noisy to allow direct visualization of vortical structures, this can be overcome by defining a quasi-condensate wavefunction $\hat{\psi}$, as established in \cite{PhysRevA.66.013603}.   This wavefunction is constructed by filtering out high-frequency spatial modes from the classical field wavefunction, by 
transforming the complex amplitudes via
$\hat{a}_{{\bf k}} = a_{{\bf k}}\times\max\{1-k^{2}/k_c^2,0\}$. $\hat{\psi}$ represents the long-wavelength component of the classical field.

The quasi-condensate features a tangle of quantized vortices which relaxes over very long times, and the final equilibrium state is free of vortices.  Its physical properties, e.g. temperature and condensate fraction, are uniquely determined by the number of particles $N$ and the kinetic energy $E=\int (\hbar^2/2m)|\nabla \psi|^2~{\rm d}V$ of the system~\cite{PhysRevLett.95.263901}.  The equilibrium state of the non-condensed particles follows the Rayleigh-Jeans distribution, modified by the nonlinear interaction with the condensed particles \cite{PhysRevLett.95.263901}.  It is interesting to note that the equilibrium condensate fraction is insensitive to the number of modes, providing that the number of modes is, or exceeds, $16^3$ modes.  This suggests that this number of modes is sufficient to model the thermodynamic limit of the system. For comparison, we employ $192^3$ modes.

Here we parametrise the system in terms of its particle density $\rho = N/\ell^3$ and average energy density $\langle H \rangle/\ell^3$.  Note that the total energy $H=E+E_0$, where $E$ is the kinetic energy of the system and $E_0$ is the energy of the condensate \cite{PhysRevLett.95.263901}. 
The temperature is evaluated from the condensate fraction using the following
empirical relationship established in Ref. \cite{berloff_2007}:
\begin{equation}
  \frac{T}{T_\lambda} = 1 - (1 - \alpha\sqrt{\rho})\frac{\rho_0}{\rho} - \alpha\sqrt{\rho}\,\left(\frac{\rho_0}{\rho}\right)^2,
  \label{eq:temp}
\end{equation}
where $T_{\lambda}$ is the critical temperature for condensation
and $\alpha=0.2275$ is a fitting parameter.
Table \ref{tbl:cond_frac} lists the parameters chosen in our simulations
and the resulting condensate fractions and temperatures
of the ensuing equilibrated classical field states.

\begin{table}
\begin{ruledtabular}
\centering
\begin{tabular}{rcccccc}
\multicolumn{7}{c}{\it Initial conditions} \\
$N/\ell^3~(\xi^{-3})$           & 0.50 & 0.50 & 0.50 & 0.50 & 0.50 & 0.50 \\
$\langle H \rangle/\ell^3~(\mu \xi^{-3})$  & 2.57 & 2.13 & 1.75 & 1.33 & 0.53 & 0.23 \\
\multicolumn{7}{c}{\it Equilibrium state} \\
$\rho_0/\rho$        & 0.02 & 0.22 & 0.36 & 0.48 & 0.77 & 0.91 \\
$T/T_\lambda$        & 0.98 & 0.81 & 0.68 & 0.56 & 0.26 & 0.10
\end{tabular}
\end{ruledtabular}
\caption{Condensate fraction and temperature of the equilibrium classical field state for our chosen initial conditions.}
\label{tbl:cond_frac}
\end{table}

\section{Moving obstacle at finite-temperature\label{sec:obstacle}}

\subsection{Critical velocity for vortex nucleation}

\begin{figure}
\centering
  \includegraphics[width=0.9\linewidth]{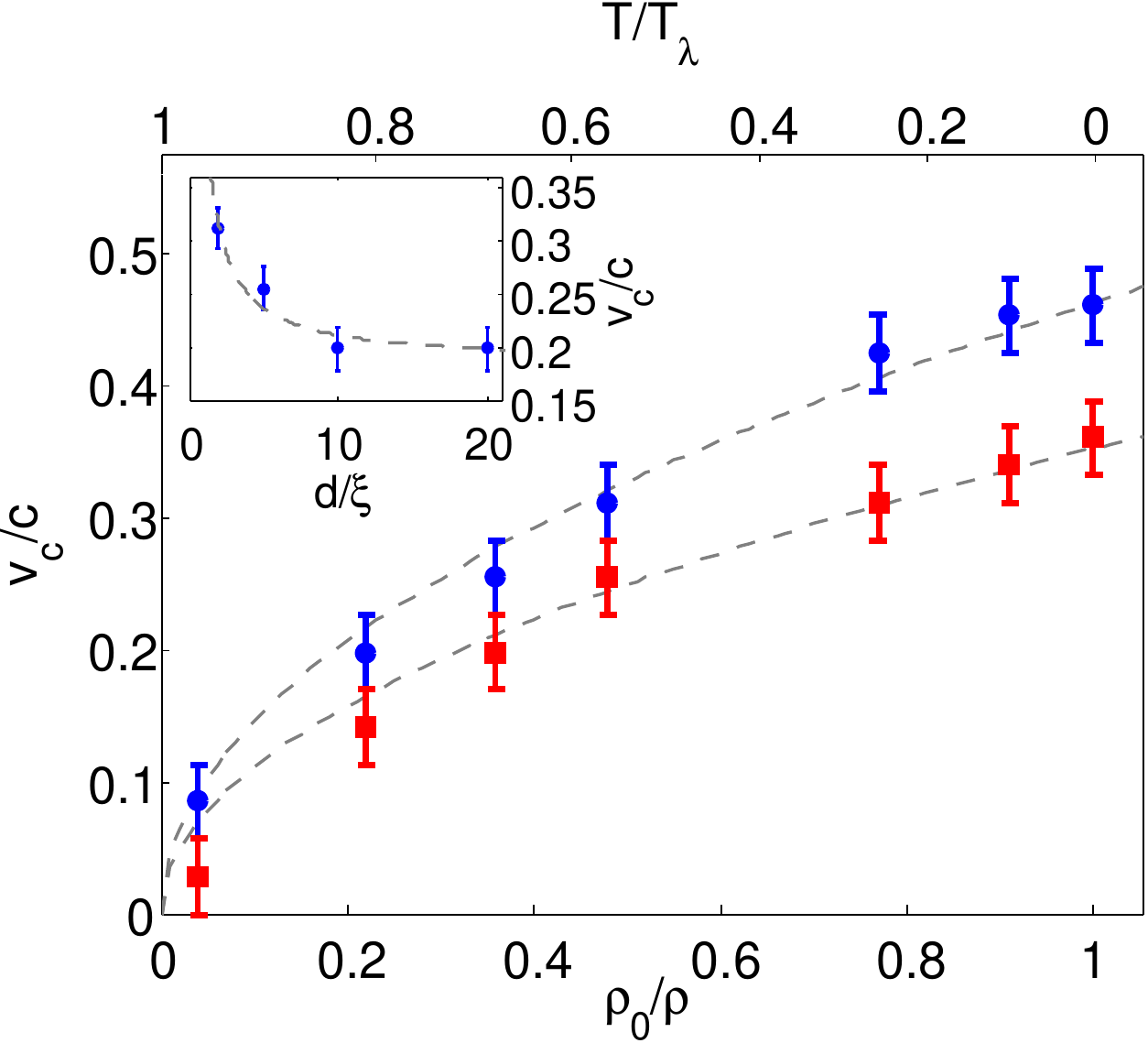}
  \caption{\label{fig:vc-n0}(Color online) Critical velocity $v_{\rm c}$ for the moving Gaussian-shaped obstacle (uniform in $z$) as a function of condensate fraction $\rho_0/\rho$ and temperature $T/T_\lambda$, for obstacle widths $d=2\xi$ (blue circles) and $d=5\xi$ (red squares). The dotted lines show the analytic model $v_{\rm c}=\beta \sqrt{\rho_0/\rho}$ with fitted coefficients $\beta=0.46$ and $0.35$. (Inset) The critical velocity approaches an asymptotic value as the obstacle size is increased. Included is a fit of the form $v_{\rm c}=\alpha/d+\gamma$ with $\alpha=0.26~(\xi^2/\tau)$ and $\gamma=0.18 c$.  Errors bars represent the systematic uncertainty in $v_c$ due to the discretized values of $v$ considered.}
\end{figure}

Having obtained the equilibrated finite-temperature states of the Bose gas, we now move on to consider a laser-induced obstacle moving through the gas.  The obstacle, uniform in $z$, is translated in the $x$-direction at speed $v$.  Our simulations are conducted in the frame moving with the obstacle, modelled by the inclusion of a Galilean shift term $i \hbar v \partial_x \psi$ to the right-hand side of the GPE.  In this frame the obstacle is imposed through the time-independent potential $V_{\mathrm{obj}}{(\bf r})=V_0 \exp\left[{-(x^2+y^2)/d^2 }\right]$, where $d$ and $V_0$ parameterize the width and amplitude of the potential.  The amplitude is linearly increased from $V_0 = 0$ at first introduction to its maximal value $V_0=5\mu$ over a period of $200\tau$.   The frame speed is increased adiabatically to the required value according to the temporal profile $v \tanh(\hat{t}/200 \tau)$, where $\hat{t}$ denotes the time from introduction of the obstacle.

Simulations are repeated (from identical initial conditions) with increasing terminal speeds (in steps of $0.057c$) until vortices are detected.  Vortex detection is by visual inspection of the filtered density, up to a maximum simulation time $\hat{t}=500\tau$ (which is long enough to ensure that the obstacle is fully introduced and at terminal speed, but otherwise arbitrary). This procedure defines the critical velocity $v_c$.  There is a systematic uncertainty in our measurement of $v_c$, arising from the discrete terminal speeds employed.  Note that we have repeated this process for multiple randomized initial conditions, and find no change in our measurement of $v_c$; that is, the systematic uncertainty due to using discretized speeds is larger than the statistical uncertainty arising from different random initial conditions.

Figure \ref{fig:vc-n0} shows the variation of $v_c$ with both condensate fraction $\rho_0/\rho$ (lower abscissa) and temperature $T/T_\lambda$ (upper abscissa), for two example obstacles widths.  The critical velocity has a maximum value at zero temperature (unit condensate fraction), and decreases nonlinearly as temperature increases (condensate fraction decreases), reaching zero at the critical point for condensation.

At zero temperature, the critical velocity is of the order of the condensate speed of sound $c=\sqrt{\rho g/m}$, with a general form $v_{\rm c}(T=0)=\beta c$,
where $\beta$ is a parameter which depends solely on the shape of the
obstacle (here $d$ and $V_0$).  The simulated $v_c$ data in
Figure \ref{fig:vc-n0} closely follow the simple functional form
$v_{\rm c}(T) = v_{\rm c}(0) \sqrt{\rho_0/\rho}$, as shown by the dashed lines.
An expression for the critical velocity valid at zero and non-zero
temperatures is
\begin{equation}
v_{\rm c}(T)=\beta \sqrt{\rho_0 g/m}.
\label{eqn:finite_vc}
\end{equation}
In other words, for a given obstacle, the critical velocity is a fixed
fraction of the speed of sound based on the {\it condensate} density
rather than the total particle density \cite{leadbeater_2003}.

The inset of Figure \ref{fig:vc-n0} shows the variation of $v_{\rm c}$ with the obstacle width $d$ at finite temperature, for the example of $T/T_\lambda =0.56$.   The qualitative behaviour is consistent with that seen at zero temperature \cite{huepe00,rica_2001,stagg_2014}: for small $d$ the critical velocity is sensitive to $d$ (due to the prominence of boundary layer effects) but as $d$ increases $v_c$ decreases towards a limiting values (the Eulerian limit).  However, the critical velocities are systematically reduced compared to the zero temperature case due to the reduced condensate speed of sound at finite temperature.

\subsection{Vortex nucleation pattern}

\begin{figure}
    \includegraphics[width=\linewidth]{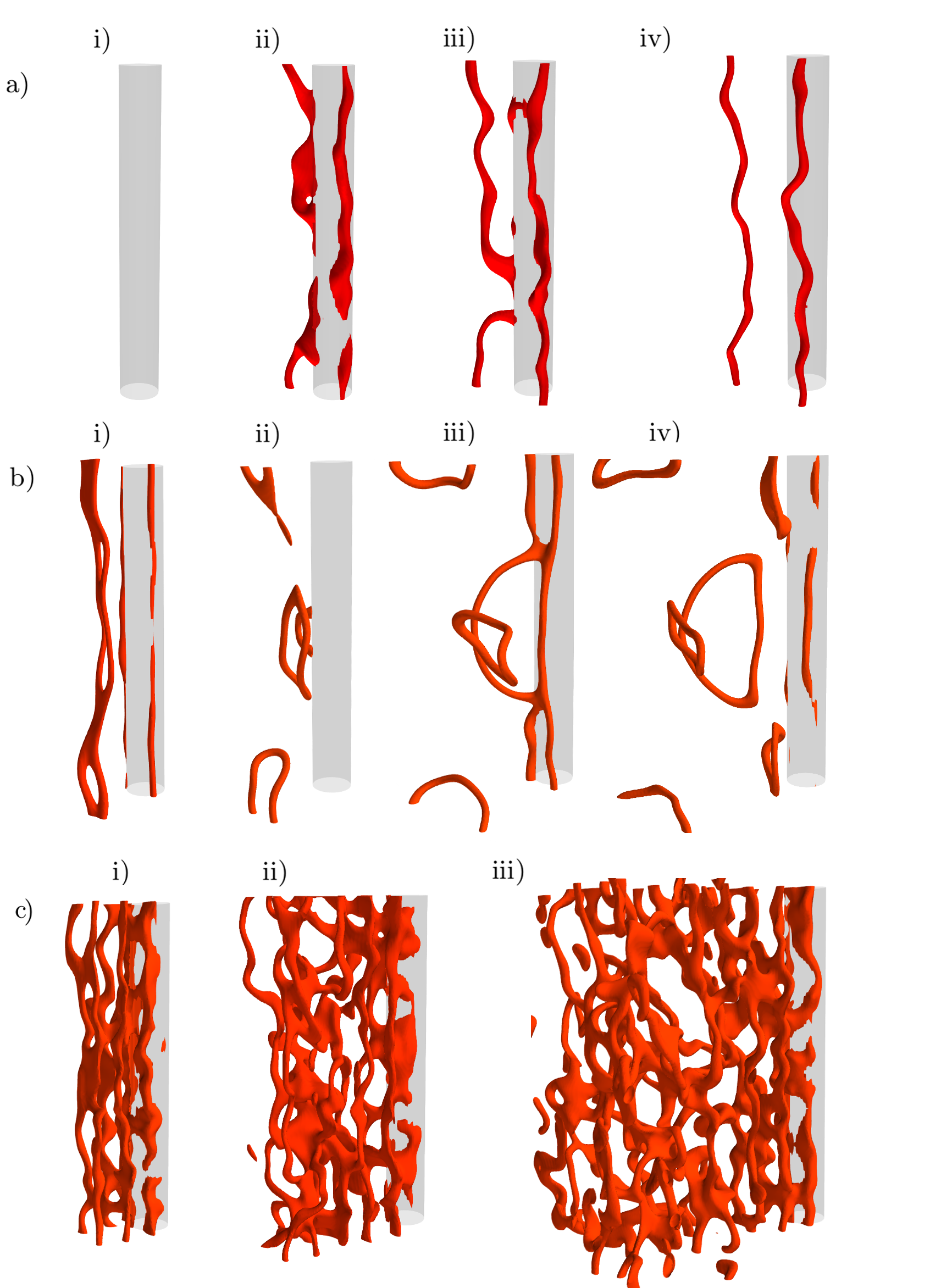}
    \caption{\label{fig:vort-lines}(Color online) Snapshots of the typical vortex nucleation from the moving Gaussian-shaped obstacle with $d=5\xi$ (gray) in the finite temperature Bose gas. The three cases (a-c) are representative of the behaviour across the whole parameter space shown in Fig. \ref{fig:vort-vals}. Shown are  isosurfaces of the quasi-condensate density ($|\hat{\psi}|^2 = 0.04\braket{|\hat{\psi}|^2}$).  (a) Vortices are shed as pairs of anti-parallel vortex lines.  Here the system parameters are $\rho_0/\rho = 0.22$ and $v=0.17c$, and the snapshots correspond to times (i) $\hat{t}/\tau=210$, (ii) $460$, (iii) $585$ and (iv) $710$.  (b) Vortex rings are nucleated from the obstacle.  The system parameters are $\rho_0/\rho = 0.91$ and $v=0.42c$, and the times are (i) $\hat{t}/\tau=500$, (ii) $700$, (iii) $875$ and (iv) $950$. (c) A vortex tangle forms behind the obstacle. The system parameters are $\rho_0/\rho = 0.35$ and $v=0.59c$, and the times are (i) $\hat{t}/\tau=250$, (ii) $375$ and (iii) $500$. }
\end{figure}

Finally we examine the manner in which vortices are nucleated from the obstacle.  At zero temperature, one expects the nucleation of straight anti-parallel vortex lines from the obstacle, either released in unison or staggered in time \cite{frisch92,saito_2010,stagg_2014}, which move downstream relative to the obstacle.  At finite temperature, we observe three general regimes of vortex nucleation, with representative examples shown in Fig.  \ref{fig:vort-lines}:

\begin{description}
\item[Vortex lines] A pair of ``wiggly'' vortex lines is produced  [Fig.  \ref{fig:vort-lines}(a)].  The wiggles are driven by the thermal fluctuations, which cause the vortex elements to be nucleated at slightly different times along the obstacle; this is visible at intermediate times (snapshots (iii) and (iv)).   These elements ultimately merge together along the axis of the obstacle to form a wiggly vortex/anti-vortex line. Similar vortex configurations
in the form of lines which are partially attached to a thin wire 
were also observed in liquid helium \cite{zieve2001}. 
\item[Vortex rings]  Here vortices predominately form vortex
rings [Figure \ref{fig:vort-lines}(b)].  The vortex loops generated
at the obstacle rapidly peel away from the obstacle, reconnecting with
adjacent loops to form rings. Due to the way the vortex rings form initially along the obstacle, they are elliptical and polarised such that they are longer along the obstacle axis. 
\item[Vortex tangle]  Strong
interaction between successively nucleated vortices leads to the formation of a complex tangle of vortex lines behind the obstacle [Figure \ref{fig:vort-lines}(c)].
\end{description}

While the vortex line regime is analogous to the zero temperature case, no analog occurs for the ring and tangle regimes.   We note that even a small amount of thermal fluctuations is enough to vastly change the form of vortex nucleation, such as the vortex rings produced in Fig.~\ref{fig:vort-lines}(b) for a condensate fraction of $0.91$.

To systematically map the occurrence of these regimes, we measure the vortex line-length density $L$ (length of vortex line per unit volume) and vortex polarity $R$ (described below) at a fixed observation time of $\hat{t}=500\tau$, throughout the parameter space of flow velocity and condensate fraction.  Our method to evaluate the vortex line-length density is described in Appendix A.  The results are presented in Fig. \ref{fig:vort-vals}.  Below the critical velocity (solid black line) no vortices are produced, and thus $L=0$.  Above the critical velocity, $L$ increases strongly with the flow velocity.  This is to be expected since the frequency of vortex nucleation increases with flow velocity \cite{frisch92}.  $L$ also increases with decreasing condensate fraction (increasing temperature), indicating the significant role of thermal fluctuations in enhancing vortex production.

Just above the critical velocity, where the vortex line-length density is relatively small, vortex nucleation occurs through vortex lines and rings.  The low flow velocity ensures that the vortex nucleation frequency is low, thereby suppressing strong interaction or reconnection between nucleated vortices.  Here, whether lines or rings are produced is sensitive to the random initial conditions, and so it is not possible to further distinguish these nucleation regimes within this parameter space. In these cases a more consistent characterisation of the vortex form is given by $R$, described below.   At higher flow velocities, where the vortex line-length density is relatively high, the nucleation frequency becomes sufficiently high that vortices immediately undergo strong interactions with each other, reconnecting and developing into a vortex tangle.  The transition in the parameter space from vortex lines/rings to tangles is indicated approximately by the dashed line, although statistical effects blur the true boundary.

We further characterise the vortex distribution by its polarisation through the quantity $R=A_z/(A_y+A_z)$, where $A_y$ and $A_z$ are the total area of vortices when the vortex distribution is projected along the $y$ and $z$ directions, respectively.  A value $R \approx 0$ corresponds to vortex lines aligned predominantly along the $z$ axis, $R\approx 1$ corresponds to lines aligned predominantly along $y$, and $R\approx 0.5$ corresponds to an isotropic vortex distribution (in the $yz$ plane).  The parameter space of $R$ has the same qualitative form as that for $L$, increasing with velocity and decreasing with condensate fraction. $R$ typically lies in the range $0.1 \salt R\salt 0.4$ for the lines/rings regime, consistent with the presence of lines which are predominantly aligned along $z$ and rings which are elongated along $z$.  It is worth noting that while the occurrence of lines or rings, for a given flow velocity and condensate fraction, is sensitive to the initial conditions, the value of $R$ is highly reproducible (to within a few percent).       For the vortex tangle regime, $0.4 \salt R\salt 0.5$.  It is worth noting that this shows that the produced tangle can be highly isotropic, despite two-dimensional nature of the obstacle that generates it.

\begin{figure}
    \includegraphics[width=\linewidth]{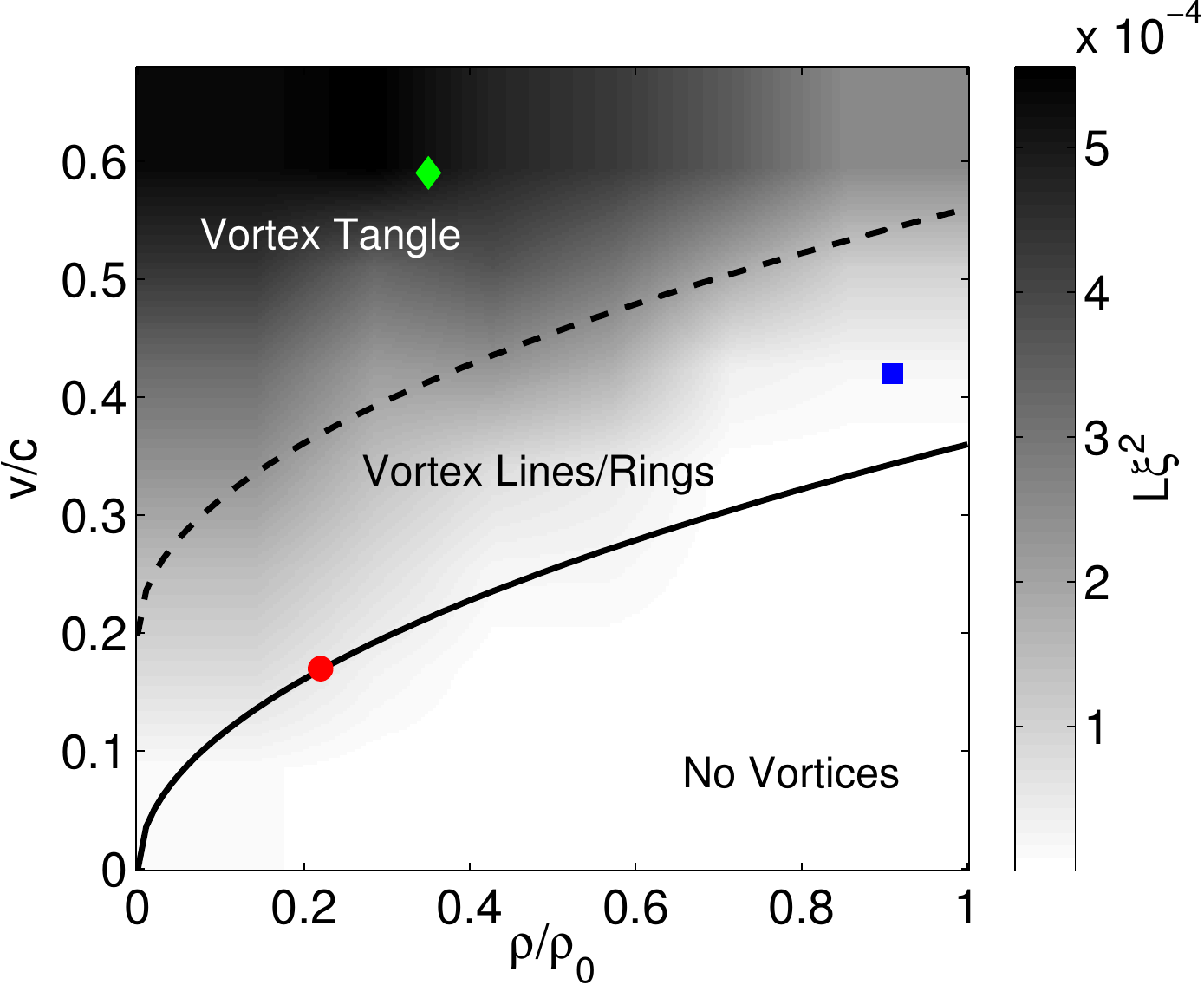}
    \caption{\label{fig:vort-vals} Vortex line density $L$ (at an observation time $\hat{t}=500\tau$)  as a function of flow speed and condensate fraction, with the qualitative regimes of vortex nucleation indicated.  The vortex line density is presented in terms of the dimensionless quantity $L \xi^2$.  The solid line marks the transition from non-vortices to vortices (i.e. the critical velocity, consistent with the corresponding fitted line in Fig. 1), while the dashed line is a guide to the eye to show the approximate transition from the vortex lines/rings regime to the vortex tangle regime.  The markers correspond to the three representative cases shown in Figure \ref{fig:vort-lines}(a) [red circle], (b) [blue square] and (c) [green diamond].  This line length density data, obtained from 36 simulations, has been interpolated.  The obstacle has size $d=5\xi$.}
\end{figure}

\section{Conclusions\label{sec:conclusions}}

Using classical field simulations, we have analysed the nucleation
of vortices past a moving cylindrical obstacle in a finite temperature
homogeneous Bose gas. We have evolved the classical field
from highly non-equilibrium initial conditions to thermalized equilibrium
states with ranging temperatures and condensate fractions.
We have then inserted
a cylindrical obstacle with Gaussian profile into the system, and imposed a flow relative to the gas.  We have found
that, above the critical velocity, vortices are nucleated forming wiggly
anti-parallel pairs of vortex lines, vortex rings, or as a vortex tangle.
The critical velocity decreases with increasing temperature,
becoming zero at the critical temperature, and scales with the speed of
sound of the condensate, i.e. as the square root of the condensate fraction.  While our work is based on a homogeneous system, in reality Bose-Einstein condensates are experimentally confined in traps, rendering the gas inhomogeneous.  Then one can expect corrections to the critical velocity due to density gradients, as well as modifications to the vortex nucleation pattern.  These higher order effects could be studied in future work.  However, we note that recent advances have led to the formation of quasi-homogeneous condensates in box-like traps \cite{gaunt_2013,chomaz_2015}, where these corrections should have minimal effect.

Data supporting this publication is openly available under an Open Data Commons Open Database License \cite{data}.

\begin{acknowledgments}
G. W. S acknowledges support from the Engineering and Physical Sciences Research Council, and N. G. P. acknowledges funding from the Engineering and Physical Sciences Research Council (Grant No. EP/M005127/1). This work made use of the facilities of N8 HPC, provided and funded by the N8 consortium and the Engineering and Physical Sciences Research Council (Grant No.EP/K000225/1). The Centre is co-ordinated by the Universities of Leeds and Manchester.
 \end{acknowledgments}
 
\appendix
\section{Evaluation of vortex line-length}
For a given wavefunction, $\psi$, featuring a vortex distribution, the vortex volume $V_t$ (the total volume associated with the vortex cores) is evaluated by numerical integration of the inside of the vortex isosurface tubes obtained from the filtered density $|\hat \psi|^2$, with an integration region of the whole numerical box.  Note that the isosurface level should be low enough to pick out vortex cores only (and not, e.g. fluctuations and waves), while large enough to contain sufficient grid points to allow a reasonable numerical evaluation of volume. In this work we use the isosurface level $0.04\langle |\hat{\psi}|^2 \rangle$ (chosen so as to produce filtered vortex cores that are similar in radius to the true vortex core).  The volume calculation can be written $V_t = \int \Theta(0.04\langle |\hat{\psi}|^2 \rangle - |\hat{\psi}({\bf r})|^2)~{\rm d}V$, where $\Theta$ is the Heaviside step function. In practice the calculation of the vortex core volume can be efficiently performed by assigning a value of unity/zero to grid points located within/outside the isosurface tubes and directly integrating the result using the trapezium or Simpson's rule.

The total line length is then deduced by dividing $V_t$ by the cross-sectional area of a vortex core, $A_t$ (in effect, the average cross-sectional area of the isosurface tubes). The measured values of $V_t$ and $A_t$ will depend on the chosen isosurface level but, providing the vortex tubes are well-separated, their ratio (and hence the evaluated line length) will remain constant.  For closely-positioned vortex tubes, the isosurface level can affect whether the tubes appear as two separate tubes, or start to merge, and so will lead to deviations in this ratio.  We have tested the effect of an alternative isosurface value.  For twice the original isosurface value, the difference in the calculated line length is negligible for systems with low vortex density.  For cases with the highest density of vortices, the difference remains less than $10\%$.


\begin{thebibliography}{10}
\bibitem{data}  DOI to be provided once accepted.
\end{thebibliography}
\end{document}